\documentclass[12pt,oneside]{article}
\usepackage{amssymb,amsmath,latexsym,amsthm,amsfonts}
\usepackage[english]{babel}
\usepackage{color,hyperref}
\usepackage[table]{xcolor} 
\usepackage{slashbox}
\usepackage{multirow}
\usepackage{graphics}
\usepackage{graphicx}
\usepackage{dsfont}
\usepackage{multicol}
\usepackage{color}
\usepackage{ulem} 
\usepackage{amsmath}

\newtheorem{prop}{Proposition}

\newtheorem{lem}{Lemma}

\numberwithin{ex}{section} \numberwithin{rem}{section}
\numberwithin{equation}{section} \numberwithin{thm}{section}
\numberwithin{lem}{section} \numberwithin{coro}{section}

\def\1g{1\hskip -3pt \mbox{l}}

\small\normalsize

\begin{document}

\begin{center}
{\Large
    {\sc  Testing for breaks in variance structures with smooth changes} }
\bigskip

Ben Hajria Raja  $^{(a)}$\footnote{Corresponding author: Faculty of Sciences of Monastir,
Laboratory of Mechanical Engineering,
University of Monastir,
5000, Monastir, Tunisia. Email: benhajriaraja@gmail.com.}, Khardani Salah   $^{(b)}$, Ra\"{i}ssi Hamdi
$^{(c)}$
\footnote{Supported by project FONDECYT 1160527.}
\bigskip

{\it

$^{(a)}$ LGM-ENIM, Faculty of Sciences of Monastir, University of Monastir,
Tunisia

$^{(b)}$ National Engineering School of Monastir,University of Monastir,
Tunisia

$^{(c)}$ Instituto de estadistica, Pontificia Universidad Catolica de
Valparaiso, Chile  }
\end{center}

\noindent {\em Abstract:} The problem of detecting variance breaks
in the case of smooth time-varying variance structure is studied. It
is highlighted that the tests based on (piecewise) constant
specification of the variance are not able to distinguish between
smooth non constant variance and the case where an abrupt change is
present. Consequently, a new procedure for detecting variance breaks
taking into account for smooth changes of the variance is proposed.
The finite sample properties of the tests introduced in the paper
are investigated by Monte Carlo experiments. The theoretical outputs
are illustrated using U.S. macroeconomic data.

\vspace*{.7cm} \noindent {\em Keywords:} Unconditionally heteroscedastic errors; Variance breaks; CUSUM test.\\

\section {Introduction}
\label{intro}

In the time series analysis literature, a considerable attention has
been paid to the test of abrupt variance breaks (see Inclan and Tiao
(1994), Berkes \textit{et al.} (2004) or Sans\'{o} \textit{et al.}
(2004) in the univariate case among others, and Aue \textit{et al.}
(2009) in the multivariate case). These tests are based on the
assumption of constant variance under the null hypothesis, which is
sometimes restrictive in the sense that continuous changes of the
variance are not taken into account.

In time series modelling it is common to reduce the time range of
the data so that the smooth variance change become negligible. For
high frequency data (daily financial data for example) it is in
general easy to define relatively large samples lengths for which the
variance could be approximated by a constant. Therefore the tools
for detecting variance breaks based on the constant variance
hypothesis under the null may be applied directly in such a case. In
such setting Berkes \textit{et al.} (2004) proposed a test to detect
abrupt changes for GARCH processes. Nevertheless for low
frequency data (for instance annual, quarterly or monthly
macroeconomic data) there are some subperiods of potential interest
for applied investigations that exhibit fast smooth changes. As a consequence such situation makes difficult to form subsamples with approximately constant
variance. In order to exemplify, let us consider the quarterly
foreign direct investment in U.S. in millions of dollars from
1946-10-01 to 2014-01-01. The series plotted in Figure \ref{i1}
shows a global increasing of the variance. Clearly if one is
interested in studying, let us say, the period beginning in the
early 90's to the end of the sample, the possible smooth changes of
the unconditional variance cannot be neglected.

The aim of this work is to investigate the test for a variance break
in presence of smooth changes. It is first established that the
tests based on the assumption of constant variance tend to reject
spuriously the hypothesis of no variance break in such a case as the
sample size increases. In practice this may lead to make a confusion
between the case where at least a variance break is present and the
case where the variance is only subject to smooth changes. As a
consequence we propose a testing procedure that is able to improve
the detection of variance breaks. Following the approach of Dahlhaus (2012, p361) the smooth changes of the variance are captured using
polynomial regressions of low orders to correct the test statistics.

The structure of the paper is as follows. In Section \ref{S2} we
show that testing for a variance break while smooth changes are
present can lead to erroneous conclusions. In Section \ref{S3} a
polynomial correction of the test statistic is proposed. In Section
\ref{S7} we carried out numerical experiments which show substantial
improvements of the control of type I errors when the polynomial
correction is applied. The outputs of the paper are illustrated
using U.S. macroeconomic data sets.

The following general notations will be used. Independently,
identically distributed is abbreviated by i.i.d..
The convergence in distribution is denoted by
$\Rightarrow$ and the symbol $\overset{p}{\to} $ denotes the
convergence in probability. If $(X_n)$ is a sequence of random
variables, then $X_n=O_p(1)$ means that $X_n$ is bounded in
probability and $X_n=o_p(1)$ means that $X_n \overset{p}{\to}  0$.
We denote by $[\cdot]$ the usual integer part of a real number. If a
lower bound of a sum exceeds the upper bound then the sum is set
equal to zero. Throughout the paper the constant $M>0$ may take
possibly different values.

\section {Unreliability of the tests based on constant variance structure}
\label{S2}

In this section it is underlined that the standard approach for
testing for a variance break may be misleading if the studied sample
(or subsample) is built so that smooth changes cannot be neglected.
For the sake of conciseness, we illustrate this only in the case
where the full sample is
considered, although similar
arguments can be used if unsuitable subsamples are taken.

Let us consider the process $(x_t)$ given by
\begin{eqnarray*}\label{VAR} &&
x_t=a_{1}x_{t-1}+\dots+a_{m}x_{t-m}+u_t\\&&
u_t=h_t\epsilon_t,\nonumber
\end{eqnarray*}
where $x_t$, $t=1,\cdots,n$ are observed random variables  and $\epsilon_t$ i.i.d. centered random variables with unit variance. It is assumed that there exists
an estimator $\hat{\theta}=(\hat{a}_{1},\dots,\hat{a}_{m})'$ for the
parameters vector $\theta_0=(a_{1},\dots,a_{m})'$ which is such that
$\sqrt{n}(\hat{\theta}-\theta_0)=O_p(1)$. For instance
$\sqrt{n}$-asymptotically normal estimators of the parameters giving
the conditional mean are provided in Xu and Phillips (2008). The
residuals are defined by
$\hat{u}_t=x_t-\hat{a}_{1}x_{t-1}-\dots-\hat{a}_{m}x_{t-m}$. Of
course if the $x_t$'s are uncorrelated, the $u_t$'s can be directly
used in the statistics introduced below.
The following assumptions delineate the framework of non constant variance structure for the errors.\\

\textbf{Assumption A1:} \quad {\bf Smooth time varying variance with
no break.}
\begin{description}
  \item[(i)] We assume that $h_t:=g(\frac{t}{n})$ where $g(\cdot)$ is a
measurable deterministic function on the interval $(0,1]$, such that
$g(r)>0$ and $\sup_{r\in(0,1]}|g(r)|<\infty$.
  \item[(ii)] The function $g(\cdot)$ satisfies a Lipschitz
condition on $(0,1]$.
  \item[(iii)] The process $(\epsilon_t)$ is such that $E(\epsilon_t^{4\delta})<\infty$ with $\delta>1$.
\end{description}
\textbf{Assumption A1': Time varying variance with a break.}\quad
Suppose that conditions (i) and (iii) of {\bf A1} are fulfilled and
that
$g(\cdot)$ is not continuous but satisfies a Lipschitz condition piecewise on two sub-intervals that partition $(0,1]$.\\

Since the rescaling device developed by Dahlhaus (1997) is used for
the definition of the $h_t$'s, $(x_t)$ should be written in a
triangular form. However the double subscript is not used to keep
the notations simple. The assumption {\bf A1} allows to consider the
realistic case where the variance evolves in a smooth way. The
assumption {\bf A1'} allows for an abrupt change for the variance in
addition to the time-varying smooth variance structure. In this
paper we develop tests to detect a variance break in a context where
smooth changes are present (i.e. $H_0$: {\bf A1} holds vs. $H_1$:
{\bf A1'} holds).


The standard situation for the null hypothesis is retrieved when
$g(.)$ is taken constant. In order to detect the presence of abrupt
breaks if the $u_t$'s are i.i.d. Gaussian, Inclan and Tiao (1994)
proposed the following statistic:

\begin{equation}\label{inclantiao}
S=\sup_k|\sqrt{n/2}\hat{D}_k|,\quad k=1,\dots,n,
\end{equation}
where $\hat{D}_k=\frac{\hat{C}_k}{\hat{C}_n}-\frac{k}{n}$,
$\hat{C}_k=\sum_{t=1}^k\hat{u}_t^2$. Sans\'{o} \textit{et al.} (2004)
proposed a corrected test statistic in the non Gaussian case:

\begin{equation}\label{sansoetal}
\widetilde{S}=\sup_k|n^{-\frac{1}{2}}\hat{B}_k|,\quad
k=1,\dots,n,\quad\mbox{with}\quad
\hat{B}_k=\frac{\hat{C}_k-\frac{k}{n}\hat{C}_n}
{\sqrt{\hat{\eta}-(n^{-1}\hat{C}_n)^2}},
\end{equation}
and $\hat{\eta}=n^{-1}\sum_{t=1}^n\hat{u}_t^4$. Under the assumption
of a constant variance and other additional conditions, it is shown
that the statistics (\ref{inclantiao}) and (\ref{sansoetal})
converge in distribution to $\sup_s|W(s)|$ where $W(s):=B(s)-sB(1)$
is a Brownian bridge, and $B(\cdot)$ being a standard Brownian motion. Of course all the results obtained in this paper for
statistics taking into account the non Gaussian case
also hold when the errors are actually independent and Gaussian
distributed.
Sans\'{o} \textit{et al.} (2004) have also proposed a statistic that can
take into account nonlinearities, which are typical in financial
data. However the non Gaussian case is adopted in the sequel
since it provides a large enough framework to handle macro-economic
data. The following proposition shows that the usual tests are not
valid in our non standard framework.

\begin{prop}\label{propostun} Under {\bf A1}, we have
$$
\widetilde{S}=o_p(n^{\frac{1}{2}})+Mn^{\frac{1}{2}},$$ where $M>0$
is a constant.
\end{prop}

From Proposition \ref{propostun} it turns out that if the smooth
changes of the variance are not taken into account correctly,
the null hypothesis of no variance break tend to be rejected spuriously by the usual tests as $n\to\infty$.\\

In order to apply the classical approach for testing for variance
breaks, usually subsamples where the variance is satisfactorily
approximated by a constant are considered. We focus on subsamples of
length $q=[n^{\gamma}]$ for some $\gamma\in(0,1)$ to illustrate this
point.
%
Let a sequence $\dot{r}_n\in(0,1)$ and introduce the following
statistic:
\begin{equation}\label{stat2}
\widetilde{S}_{\dot{r}_n}^{\gamma}=\sup_k|q^{-\frac{1}{2}}\hat{B}_{k,\dot{r}_n}^{\gamma}|,\quad
k=1,\dots,q,\quad\mbox{with}\quad
\hat{B}_{k,\dot{r}_n}^{\gamma}=\frac{\hat{C}_{k,\dot{r}_n}^{\gamma}-
\frac{k}{q}\hat{C}_{q,\dot{r}_n}^{\gamma}}
{\sqrt{\hat{\eta}_{\dot{r}_n}^{\gamma}-(q^{-1}\hat{C}_{q,\dot{r}_n}^{\gamma})^2}},
\end{equation}
with
$\hat{C}_{k,\dot{r}_n}^{\gamma}=\sum_{t=[\dot{r}_nn]+1}^{[\dot{r}_nn]+k}\hat{u}_t^2$
and
$\hat{\eta}_{\dot{r}_n}^{\gamma}=q^{-1}\sum_{t=[\dot{r}_nn]+1}^{[\dot{r}_nn]+q}\hat{u}_t^4$.
Therefore the $\widetilde{S}_{\dot{r}_n}^{\gamma}$ statistic is computed at
fractions $\dot{r}_n$ of the original sample with a subsample of
length $q$. Note that $\dot{r}_n$ should be chosen adequately in
view of the sample size, $[\dot{r}_nn]+q<n$.
For mathematical convenience the increasing sequence
$\dot{r}_n$ is such that the subsample middle $r^0$ is fixed. Also it is assumed that a possible variance break necessarily occurs in $r^0$.
Note that the above setting can be replaced by the assumption that $\dot{r}_n$ is increasing, so that the abrupt change is
present in all subsamples as $q\to\infty$ for power results.
The terms $\gamma$ and
$\dot{r}_n$ may be viewed as parameters for calibrating the
subsamples of interest.
The following proposition gives the
asymptotic behavior of the $\widetilde{S}_{\dot{r}_n}^{\gamma}$
statistic.

\begin{prop}\label{modtest1} Suppose that $0<\gamma\leq \frac{2}{3}$. Then under {\bf A1} we have as $q\to\infty$,
$
\widetilde{S}_{\dot{r}_n}^{\gamma}\Rightarrow\sup_{s\in(0,1]}|W(s)|.$
\end{prop}

The proof of Proposition \ref{modtest1} is given in the Appendix.
The following result ensures the
consistency of the test based on the standard statistic and subsamples where the variance can be approximated by a constant.

\begin{prop}\label{modtest2} Under \textbf{A1'} we have $\widetilde{S}_{\dot{r}_n}^{\gamma}=Mn^{\frac{\gamma}{2}}+O_p(1)$
, where $M>0$ is a constant.
\end{prop}The above results give a testing procedure which corresponds to the
common practice consisting in selecting a subsample where the smooth
changes in the variance structure can be neglected, so that the
classical tests may be applied directly. Indeed it is well known
that the processes given by assumption {\bf A1} can be viewed as
approximately stationary (see e.g. Dahlhaus (2012)).

In general it is clear that marked smooth changes may lead to select
too small subsamples with almost constant variance under the null of no
variance breaks. Indeed, although Proposition \ref{modtest1} and
\ref{modtest2} ensure a good implementation of the classical tests
as $n\to\infty$ for suitable subsamples, the lengths of low
frequency economic series are too small in many cases. Hence the
detection of variance breaks may become intractable and could lead
to size distortions problems.
On the other hand the approximate constant variance may be
questionable when too large subsamples are selected, so that we can
loose the control of the type I error in view of Proposition
\ref{propostun}. Note also that the practitioner may be interested
in analyzing the data on larger samples than those that allow to
neglect the smooth variance changes.
In the next section a procedure for testing a variance break in presence of marked smooth changes is proposed.\\

\section {Testing for variance break handling smooth changes in the variance structure}
\label{S3}

Assume that under \textbf{A1} we can write
\begin{equation*}\label{regression poly}
g^2(r)=\sum_{i=0}^p{\alpha_{i,r^0}(r-r^0)^i+o((r-r^0)^p)},
\end{equation*}
for some $p>0$ and for any $r$, $r^0\in(0,1)$. In the same way as
before a subsample of length $q=[n^\gamma]$, is taken. For a
potentially better precision, we use
$r^0:=(2\frac{[\dot{r}_nn]}{n}+\frac{q}{n})/2$, the subsample
middle, and the coefficients are estimated by ordinary least squares
(OLS) from the following equation:
\begin{equation}\label{regression1}
u_t^2=\sum_{i=0}^p{\alpha_{i,r^0}\left(\frac{t}{n}-r^0\right)^i}+\xi_t,
\end{equation}
where $\xi_t=u_t^2-g^2\left(\frac{t}{n}\right)$ is the error term and $t=[\dot{r}_nn]+1,\dots,[\dot{r}_nn]+q$.
As a reduced subsample size is considered,
we can think that a relatively small order $p$ describes
satisfactorily the smooth time varying variance structure. Let
$\hat{\alpha}_{i,r^0}$ denote the (OLS) estimators and
$\hat{g}^2(r)$ the estimated variance. It is shown in Lemma
\ref{lemma 1}  that $\hat{\alpha}_{i,r^0}$ is a consistent
estimator of $\alpha_{i,r^0}$, so that a smooth approximation of
the variance structure is available. Suppose that $g^2(r)>c>0$,
which implies that $\hat{g}^2(r)>c>0$ for large enough $q$. Define
the test statistic:
\begin{equation}\label{stat3}
\bar{S}_{\dot{r}_n}^{\gamma}=\sup_k|q^{-\frac{1}{2}}\bar{B}_{k,\dot{r}_n}^{\gamma}|,\quad
k=1,\dots,q,\quad\mbox{with}\quad
\bar{B}_{k,\dot{r}_n}^{\gamma}=\frac{\bar{C}_{k,\dot{r}_n}^{\gamma}-
\frac{k}{q}\bar{C}_{q,\dot{r}_n}^{\gamma}}
{\sqrt{\bar{\eta}_{\dot{r}_n}^{\gamma}-(q^{-1}\bar{C}_{q,\dot{r}_n}^{\gamma})^2}},
\end{equation} and
$\bar{\eta}_{\dot{r}_n}^{\gamma}=q^{-1}\sum_{t=[\dot{r}_nn]+1}^{[\dot{r}_nn]+q}\hat{g}^{-4}(\frac{t}{n})\hat{u}_t^4$,
$\bar{C}_{k,\dot{r}_n}^{\gamma}=\sum_{t=[\dot{r}_nn]+1}^{[\dot{r}_nn]+k}\hat{g}^{-2}(\frac{t}{n})\hat{u}_t^2$.
Thus we propose to use a statistic corrected from the smooth changes
of the variance under the null hypothesis. For $p=0$, it is better
to use the simple tests described in the previous section. The
following propositions give the asymptotic behavior of the statistic
$\bar{S}_{\dot{r}_n}^{\gamma}$.

\begin{prop}\label{propo4} Suppose that {\bf A1} holds true, then as
$q\to\infty$,
\begin{equation*}
\bar{S}_{\dot{r}_n}^{\gamma}\Rightarrow\sup_{s\in(0,1]}|W(s)|.
\end{equation*}
\end{prop}

\begin{prop}\label{propo5} Suppose that {\bf A1'} holds true, then as $q\to\infty$,
$\bar{S}_{\dot{r}_n}^{\gamma}=Mn^{\frac{\gamma}{2}}+O_p(1)$, where
$M>0$ is a constant.
\end{prop}

Using Proposition \ref{propo4} and \ref{propo5}, we can construct a
valid test to detect variance breaks taking into account the smooth
changes of the variance.
For a suitable polynomial of order $p$ the test consists in rejecting
the null hypothesis at the asymptotic level 5\% if the test
statistic $\bar{S}_{\dot{r}_n}^{\gamma}$ exceeds the usual critical
value of the supremum of a standard Brownian bridge.

\section {Monte Carlo experiments}
\label{S7}

In the sequel, we denote by $Q_{mod}$ the modified test subject to polynomial
regression correction and with  polynomial order selection by AIC
criteria. The standard test proposed in Sans\'{o} et al.
(2004) is denoted by $Q_{std}$. In this section the finite sample properties of the $Q_{mod}$ and
$Q_{std}$ tests is examined by
simulations. We consider two data generating processes :
\begin{eqnarray}\label{DGP}
\begin{array}{c l l l l}
    DGP\,1 &:& u_t & = & h_t\epsilon_t, \\
    DGP\,2 &:& x_t & = & 0.4 x_{t-1}+u_t  \\
    &&u_t&=&h_t\epsilon_t,
   \end{array}
\end{eqnarray} where the process $\epsilon_t$ is i.i.d. and follows
the standard logistic distribution. In DGP1 the $u_t$'s are directly
observed. The autoregressive parameter in DGP2 is estimated by OLS.  The residuals are then used to
build the different statistics. Note that the errors $(u_t)$ have
non constant unconditional variance if the $h_t$'s change over time.
We carried out experiments with different settings for the variance
structure. An extract which reflects the outputs we obtained is
provided. We consider:
\begin{eqnarray}\label{fonction}
h^2(t)=-2.7+1.5\,\exp\,\left(1+\left(\frac{t}{n}\right)\right)+0.2\,\sin\,\left(5\,\pi\left(\frac{t}{n}\right)\right)+f(t),
\end{eqnarray}
with $$f(t)=\alpha
\mathds{1}_{\{t\geq[n\kappa]\}}, \quad \kappa=0.5,
\quad t=1,\cdots,n\quad and\quad \alpha=0,\;1,\: 2,\:\cdots,\: 5.$$
The first term in (\ref{fonction}) gives a global increasing
behavior for the variance structure, while the second describes a
cyclical behavior often observed in practice. The term $\alpha$ is used for the empirical power study. For each experiment
$N=1000$ independents trajectories are simulated using DGP1 and
DGP2. Samples of length $n=50$, $n=100$ and $n=200$ are simulated. In all our experiments the level of the tests is $5\%$.

\subsection {The behavior of the studied tests under the null hypothesis}

We study the empirical size of the tests, that is testing for a variance
break in presence of smooth changes. To this aim we
set $\alpha=0$ in (\ref{fonction}). The results are provided in
Tables \ref{table1} and \ref{table1e}. Assuming that the finite
sample size of the test is $5\%$, and noting that $N=1000$
replications are performed, the relative rejection frequencies
should be between the significant limits $3.65\%$ and $6.35\%$ with
probability $0.95$. Tables \ref{table1} and \ref{table1e} reveal
that, when the unconditional variance is not constant, the standard
test spuriously rejects the null hypothesis as the sample size
becomes large. On the other hand, it can be seen that the $Q_{mod}$
test improves substantially the control of the type I errors.
\begin{table}[!h]
\begin{center}
\begin{tabular}{|c|c|c|c|}
  \hline
\backslashbox {Test Statistics}{n} &\textbf{50} & \textbf{100} & \textbf{200} \\
  \hline
   \hline
 $Q_{std}$ &24.0&57.4&90.2\\
     \hline
  $Q_{mod}$ &1.0&2.9&5.5\\
  \hline
\end{tabular}
\end{center}
\caption{Empirical size (in \%) of the tests under
DGP1.\label{table1}}
\end{table}

\begin{table}[!h]
\begin{center}
\begin{tabular}{|c|c|c|c|}
  \hline
\backslashbox {Test Statistics}{n} &\textbf{50} & \textbf{100} & \textbf{200} \\
  \hline
   \hline
 $Q_{std}$ &45.2&88.1&99.6\\
     \hline
  $Q_{mod}$ &0.9&3.1&4.9 \\
  \hline
\end{tabular}
\end{center}
\caption{Empirical size (in \%) of the tests under
DGP2.\label{table1e}}
\end{table}

\subsection {The behavior under the alternative hypothesis}

In the empirical power of this section, we examine the ability of
$Q_{mod}$ test to detect an abrupt volatility break. We simulate
$N=1000$ independent trajectories using the data generating processes presented in (\ref{DGP}) with
break at level $\kappa=0.5$, taking $\alpha=1,2,3,4,5$ in
(\ref{fonction}). Tables \ref{table2} and \ref{table2e} show the
empirical powers of the $Q_{mod}$ test. As expected, the rejections
rates increase as $\alpha$ and $n$ are increased. Nevertheless we
note a low power, although the $Q_{mod}$ have some ability to detect
breaks. This is the price to pay for controlling the type I errors.

\begin{table}[!h]
\begin{center}
\begin{tabular}{|c|c|c|c|}
  \hline
\backslashbox {Break length}{n} &\textbf{50} & \textbf{100} & \textbf{200} \\
  \hline
   \hline
  $\alpha=1$ &2.0&4.8&7.1 \\
     \hline
  $\alpha=2$ &3.2&6.6&9.9 \\
     \hline
  $\alpha=3$ &2.2&6.8&13.6 \\
     \hline
  $\alpha=4$ &3.1&7.4&17.5 \\
     \hline
  $\alpha=5$ &4.1&10.0&19.7 \\
  \hline
\end{tabular}
\end{center}
\caption{Empirical power (in \%) of the $Q_{mod}$ test under
DGP1.\label{table2}}
\end{table}

\begin{table}[!h]
\begin{center}
\begin{tabular}{|c|c|c|c|c|c|c|c|c|c|c|}
  \hline
\backslashbox {Break length}{n} &\textbf{50} & \textbf{100} & \textbf{200}\\
  \hline
   \hline
  $\alpha=1$ &2.0&3.5&7.2\\
     \hline
  $\alpha=2$ &2.4&5.5&10.0\\
     \hline
  $\alpha=3$ &2.1&6.0&14.0\\
     \hline
  $\alpha=4$ &3.2&6.0&18.0 \\
     \hline
  $\alpha=5$ &3.4&9.0&19.4 \\
  \hline
\end{tabular}
\end{center}
\caption{Empirical power (in \%) of the $Q_{mod}$ test under
DGP2.\label{table2e}}
\end{table}

\section {Illustrative examples}

\begin{figure}
\begin{minipage}[c]{.46\linewidth}
\includegraphics[scale=0.5]{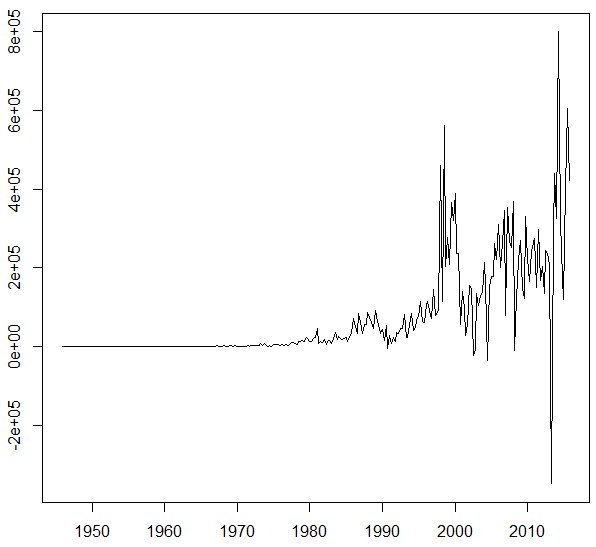}
\end{minipage} \hfill
\begin{minipage}[c]{.46\linewidth}
\includegraphics[scale=0.5]{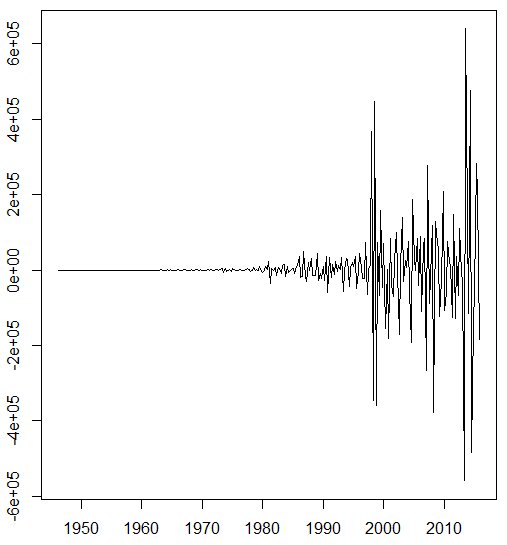}
 \end{minipage}
 \caption{The quarterly foreign direct investment in U.S. in
millions of dollars from 1946-10-01 to 2014-01-01 (n= 250) on the left, and their first
differences on the right. Data source: The research division of the
federal reserve bank of Saint Louis, code ROWFDIQ027S,
www.research.stlouis.org.} \label{i1}
\end{figure}

\begin{figure}
\begin{minipage}[c]{.46\linewidth}
\includegraphics[scale=0.55]{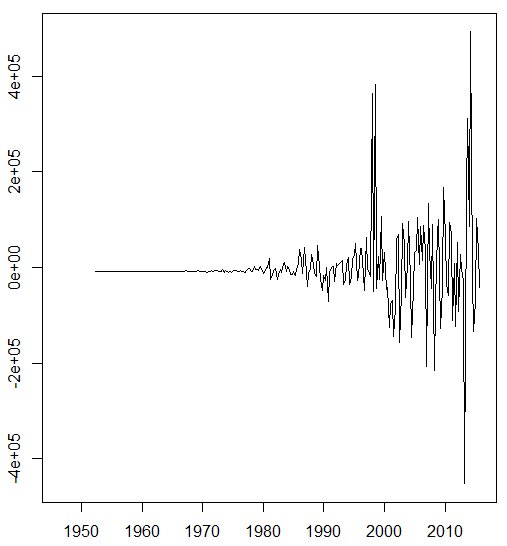}
\end{minipage} \hfill
\begin{minipage}[c]{.45\linewidth}
\includegraphics[scale=0.5]{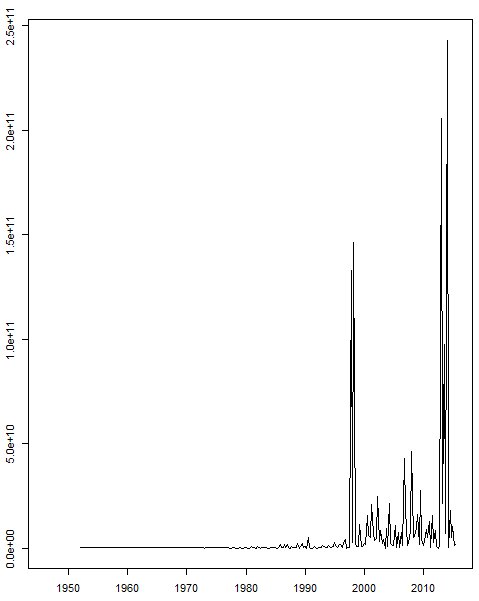}
 \end{minipage}
\caption{The OLS residuals for the foreign direct
investment data on the left, and their squares on the right.} \label{a1}
\end{figure}

\begin{figure}
\begin{minipage}[c]{.46\linewidth}
\includegraphics[scale=0.62]{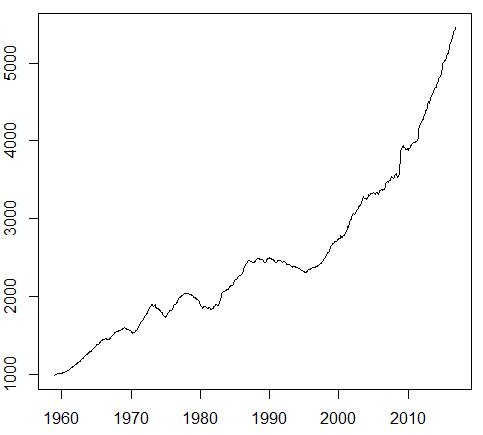}
\end{minipage} \hfill
\begin{minipage}[c]{.46\linewidth}
\includegraphics[scale=0.45]{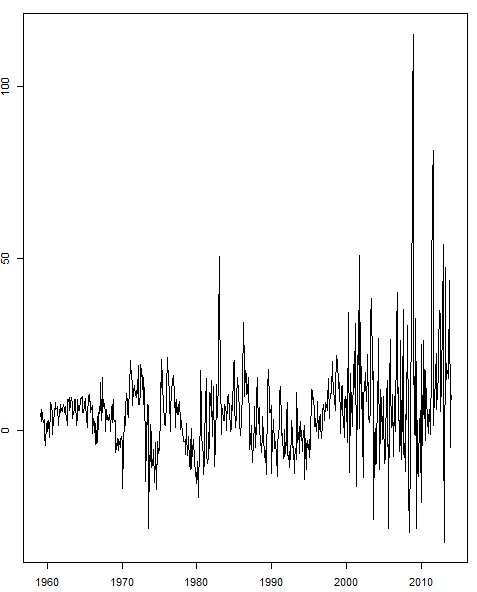}
 \end{minipage}
\caption{The monthly real M2 money stock  in U.S. in
billions of dollars
from 1959-01-01 to 2014-01-01 (n= 694) on the left,
and their first differences on the right. Data source: The research division
of the federal reserve bank of Saint Louis, code: M2REAL,
www.research.stlouis.org.} \label{i2}
\end{figure}

\begin{figure}
\begin{minipage}[c]{.46\linewidth}
\includegraphics[scale=0.55]{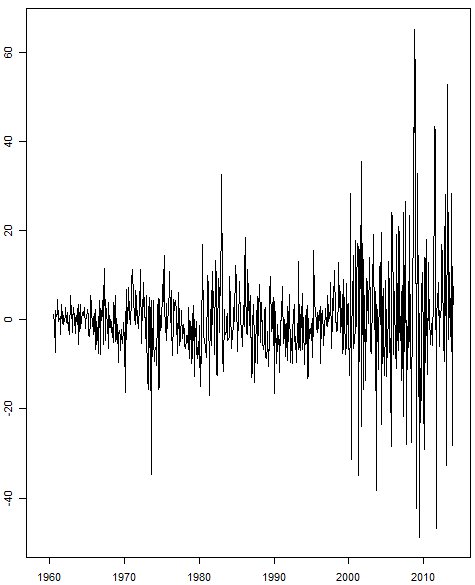}
\end{minipage} \hfill
\begin{minipage}[c]{.46\linewidth}
\includegraphics[scale=0.55]{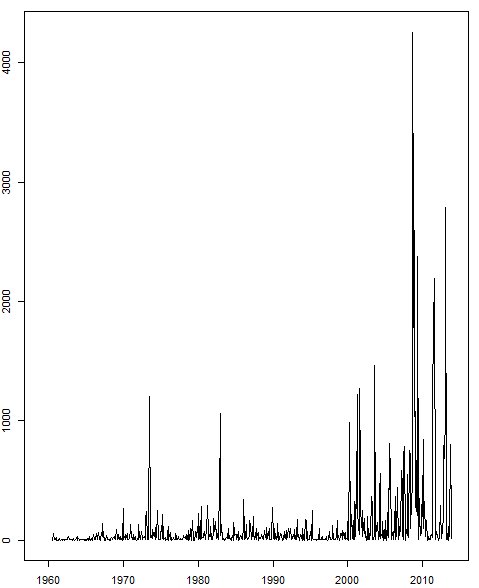}
 \end{minipage}
\caption{The OLS residuals of the real M2 money
stock data on the left, and their squares on the right.}
\label{a2}
\end{figure}
Now we turn to several applications of the test developed above to
real data sets for which it is reasonable to suppose at least smooth non constant
variance. The standard test is also used for comparison. We
investigate two macroeconomic data sets: the first differences of
the monthly real M2 money stock in billions of dollars
(hereafter noted M2) and the first difference series of the quarterly foreign direct
investment in the U.S. in millions of dollars from October 1946 to
January 2014 (called FDI hereafter). The two studied
series are plotted in Figures (\ref{i1}) and (\ref{i2}).
The data are available seasonally adjusted from the website of the
research division of the federal reserve bank of Saint Louis
(www.research.stlouisfed.org). Note that such series are often
included in many applied works.

In order to study the variance structure of residuals, we fitted AR
models to the M2 and FDI series. It appears reasonable to
assume that the variance of the residuals of these series is not
constant in time, but rather have an increasing behavior. We aim to test if in addition to smooth time varying behaviors, abrupt breaks are present. The
$Q_{mod}$ test is then applied to the residuals of the M2 and
FDI series. The outputs are compared with those of the standard test
in table \ref{tab3}. We first remark that the $Q_{std}$ test
statistic exceeds the predetermined boundary 1,33 which corresponds
to the asymptotic critical values of the supremum of a standard
Brownian Bridge (see table 1 of Sans\'{o} et al. (2004)) in all
investigated cases. As a consequence the presence of a break in the
variance structure is detected using the standard test. In view of
our results, this is possibly due to neglected smooth time-varying
variance. Now eliminating the effect of possible smooth changes, it
appears that for the M2 serie the value of $Q_{mod}$ is
lower than the asymptotic value 1.33 so that the null hypothesis of
no variance break cannot be rejected. On the other hand we can see
that the $Q_{mod}$ exceeds the predetermined boundary for the FDI.
The results in table \ref{tab3} reveal that the outputs for the
$Q_{std}$ and $Q_{mod}$ are quite different.

\begin{table}[!h]
\begin{center}
\begin{tabular}{|c|c|c|c|}
  \hline
    &$Q_{std}$&$Q_{mod}$&AIC-Order\\
  \hline
  \hline
M2 &4.14&1.26&3 \\
  \hline
FDI &2.39&1.45&3 \\
  \hline
\end{tabular}
\end{center}
\caption{ The $Q_{mod}$ and $Q_{std}$ tests based on residuals from
the first-difference of foreign direct investment and real M2 money
stock in U.S. series.\label{tab3}}
\end{table}

\section* {Proofs}
Recall that we defined
$\hat{u}_t=x_t-\hat{a}_{1}x_{t-1}-\dots-\hat{a}_{m}x_{t-m}$ the
residuals obtained from $\hat{\theta}$. From the Mean Value Theorem
it is easy to see that
$n^{-\frac{1}{2}}\sum_{t=1}^n\hat{u}_t^2=n^{-\frac{1}{2}}\sum_{t=1}^nu_t^2+o_p(1)$,
and hence the possibly unobserved process $(u_t)$ will be used for
our asymptotic derivations without loss of generality. Define
$C_k=\sum_{t=1}^ku_t^2$
$B_k=\frac{C_k-\frac{k}{n}C_n}{\sqrt{\eta-(n^{-1}C_n)^2}},$ and
$\eta=n^{-1}\sum_{t=1}^nu_t^4$.
Recall also that the general constant $M>0$ may take different values.\\

\noindent{\bf Proof of Proposition \ref{propostun}.}\quad First
using Phillips and Xu (2006), Lemma 1, we write for any $s\in(0,1]$
\begin{eqnarray}\label{ph2}
n^{-1}\sum_{t=1}^{[ns]}u_t^2=\int_0^sg^2(r)dr+o_p(1), \:\mbox{and}\:
n^{-1}\sum_{t=1}^{n}u_t^4=E(\epsilon_1^4)\int_0^1g^4(r)dr+o_p(1).
\end{eqnarray}
Noting that
\begin{eqnarray*}
|n^{-\frac{1}{2}}B_{[ns]}|&=&\left|\frac{n^{-\frac{1}{2}}(C_{[ns]}-\frac{[ns]}{n}C_n)}{\sqrt{\eta-(n^{-1}C_n})^2}\right|\\
&=&n^{-\frac{1}{2}}\left|C_{[ns]}-\frac{[ns]}{n}C_n\right|\times\left[\eta-(n^{-1}C_n)^2\right]^{-\frac{1}{2}}\\
&=&n^{-\frac{1}{2}}\left|\sum_{t=1}^{[ns]}u_t^2-\frac{[ns]}{n}\sum_{t=1}^{n}u_t^2\right|\times\left[n^{-1}\sum_{t=1}^{n}u_t^4-\left(n^{-1}\sum_{t=1}^{n}u_t^2\right)^2\right]^{-\frac{1}{2}},
\end{eqnarray*}
from (\ref{ph2}), we obtain
\begin{eqnarray}\label{HS}
|n^{-\frac{1}{2}}B_{[ns]}|&=&n^{\frac{1}{2}}\left|\int_0^sg^2(r)dr
                    -s\int_0^1g^2(r)dr\right|\nonumber\\&\times&\left[E(\epsilon_1^4)\int_0^1g^4(r)dr-\left(\int_0^1g^2(r)dr\right)^2\right]^{-\frac{1}{2}}+o_p(\sqrt{n}),
\end{eqnarray}
For the first term on the right hand side of (\ref{HS}) we have
\begin{eqnarray*}
\sup_{s\in (0,1]} \left|\int_0^sg^2(r)dr-s\int_0^1g^2(r)dr\right|>0
\end{eqnarray*}
provided that $g(\cdot)$ is not constant, while the second term is
clearly equal to a strictly positive constant. Hence we obtain
$$\sup_k|n^{-\frac{1}{2}}\hat{B}_k|=n^{\frac{1}{2}}M+o_p(\sqrt{n}),$$ which proves Proposition \ref{propostun}.$\quad\square$\\

\noindent{\bf Proof of Proposition \ref{modtest1}.}\quad We compare
the statistic defined by (\ref{stat2}) to the statistic calculated
from a subsample based on the constant variance assumption, defined
as
\begin{eqnarray*}\label{stat4}
\dot{S}_{\dot{r}_n}^{\gamma}=\sup_k|q^{-\frac{1}{2}}\dot{B}_{k,\dot{r}_n}^{\gamma}|,\quad\mbox{with}\quad
\dot{B}_{k,\dot{r}_n}^{\gamma}=\frac{\dot{C}_{k,\dot{r}_n}^{\gamma}-
\frac{k}{q}\dot{C}_{q,\dot{r}_n}^{\gamma}}
{\sqrt{\dot{\eta}_{\dot{r}_n}^{\gamma}-(q^{-1}\dot{C}_{q,\dot{r}_n}^{\gamma})^2}},\quad
k=1,\dots,q,
\end{eqnarray*}where $\dot{C}_{k,\dot{r}_n}^{\gamma}=\sum_{t=[\dot{r}_nn]+1}^{[\dot{r}_nn]+k}g^2(\frac{[\dot{r}_nn]}{n})\epsilon_t^2$
and
$\dot{\eta}_{\dot{r}_n}^{\gamma}=q^{-1}\sum_{t=[\dot{r}_nn]+1}^{[\dot{r}_nn]+q}g^4(\frac{[\dot{r}_nn]}{n})\epsilon_t^4$.\\
There are two parts of the proof of proposition 2, we study the
nominator and the denominator in (\ref{stat2}) separately. For the
nominator, we have
\begin{eqnarray*}
&&\left|\frac{1}{\sqrt{q}}\left(C_{k,\dot{r}_n}^{\gamma}-\frac{k}{q}C_{q,\dot{r}_n}^{\gamma}\right)-\frac{1}{\sqrt{q}}\left(\dot{C}_{k,\dot{r}_n}^{\gamma}-\frac{k}{q}\dot{C}_{q,\dot{r}_n}^{\gamma}\right)\right|\\
&=&\frac{1}{\sqrt{q}}\left|\sum_{t=[\dot{r}_nn]+1}^{[\dot{r}_nn]+k}\left[g^2\left(\frac{t}{n}\right)-g^2\left(\frac{[\dot{r}_nn]}{n}\right)\right]\epsilon_t^2-\frac{k}{q}\sum_{t=[\dot{r}_nn]+1}^{[\dot{r}_nn]+q}\left[g^2\left(\frac{t}{n}\right)-g^2\left(\frac{[\dot{r}_nn]}{n}\right)\right]\epsilon_t^2\right|\\
&\leq&\frac{1}{\sqrt{q}}\left|\sum_{t=[\dot{r}_nn]+1}^{[\dot{r}_nn]+k}\left[g^2\left(\frac{t}{n}\right)-g^2\left(\frac{[\dot{r}_nn]}{n}\right)\right]\epsilon_t^2\right|+\frac{k}{q^{\frac{3}{2}}}\left|\sum_{t=[\dot{r}_nn]+1}^{[\dot{r}_nn]+q}\left[g^2\left(\frac{t}{n}\right)-g^2\left(\frac{[\dot{r}_nn]}{n}\right)\right]\epsilon_t^2\right|\\
&\leq&\sup_{[\dot{r}_nn]+1\leq t\leq[\dot{r}_nn]+q}\left|g^2\left(\frac{t}{n}\right)-g^2\left(\frac{[\dot{r}_nn]}{n}\right)\right|\left(\frac{1}{\sqrt{q}}\left|\sum_{t=[\dot{r}_nn]+1}^{[\dot{r}_nn]+k}\epsilon_t^2\right|+\frac{k}{q^{\frac{3}{2}}}\left|\sum_{t=[\dot{r}_nn]+1}^{[\dot{r}_nn]+q}\epsilon_t^2\right|\right)\\
&\leq&Mq^{1-\frac{1}{\gamma}}\left(\frac{1}{\sqrt{q}}\left|\sum_{t=[\dot{r}_nn]+1}^{[\dot{r}_nn]+k}(\epsilon_t^2-1)\right|+\frac{k}{q^{\frac{3}{2}}}\left|\sum_{t=[\dot{r}_nn]+1}^{[\dot{r}_nn]+q}(\epsilon_t^2-1)\right|+\frac{2k}{\sqrt{q}}\right),
\end{eqnarray*}where the last inequality follows from the Lipschitz condition, then it follows from the Donsker's functional central limit theorem that for all $0<\gamma<\frac{2}{3}$,
\begin{eqnarray}\label{RS}
\left|\frac{1}{\sqrt{q}}\left(C_{k,\dot{r}_n}^{\gamma}-\frac{k}{q}C_{q,\dot{r}_n}^{\gamma}\right)-\frac{1}{\sqrt{q}}\left(\dot{C}_{k,\dot{r}_n}^{\gamma}-\frac{k}{q}\dot{C}_{q,\dot{r}_n}^{\gamma}\right)\right|=o_p(1).
\end{eqnarray} For the denominator we introduce
\begin{eqnarray*}
\tau^2=\hat{\eta}_{\dot{r}_n}^{\gamma}-(q^{-1}\hat{C}_{q,\dot{r}_n}^{\gamma})^2=\frac{1}{q}\sum_{t=[\dot{r}_nn]+1}^{[\dot{r}_nn]+q}g^4\left(\frac{t}{n}\right)\epsilon_t^4-\left[
\frac{1}{q}\sum_{t=[\dot{r}_nn]+1}^{[\dot{r}_nn]+q}g^2\left(\frac{t}{n}\right)\epsilon_t^2\right]^2
\end{eqnarray*}
and
\begin{eqnarray*}
\dot{\tau}^2=\dot{\eta}_{\dot{r}_n}^{\gamma}-(q^{-1}\dot{C}_{q,\dot{r}_n}^{\gamma})^2=\frac{1}{q}\sum_{t=[\dot{r}_nn]+1}^{[\dot{r}_nn]+q}g^4\left(\frac{[\dot{r}_nn]}{n}\right)\epsilon_t^4-\left[
\frac{1}{q}\sum_{t=[\dot{r}_nn]+1}^{[\dot{r}_nn]+q}g^2\left(\frac{[\dot{r}_nn]}{n}\right)\epsilon_t^2\right]^2.
\end{eqnarray*}Using the Lipschitz condition and the law of large numbers, we
obtain
\begin{eqnarray}\label{tau1}
\left|\frac{1}{q}\sum_{t=[\dot{r}_nn]+1}^{[\dot{r}_nn]+q}\left[g^2\left(\frac{t}{n}\right)-g^2\left(\frac{[\dot{r}_nn]}{n}\right)\right]\epsilon_t^2\right|\leq
Mq^{\frac{(\gamma-1)}{\gamma}}\left|\frac{1}{q}\sum_{t=[\dot{r}_nn]+1}^{[\dot{r}_nn]+q}\epsilon_t^2\right|=o_p(1).
\end{eqnarray}Similarly, we can write
\begin{eqnarray}\label{tau2}
\left|\frac{1}{q}\sum_{t=[\dot{r}_nn]+1}^{[\dot{r}_nn]+q}g^4\left(\frac{t}{n}\right)\epsilon_t^4-\frac{1}{q}\sum_{t=[\dot{r}_nn]+1}^{[\dot{r}_nn]+q}g^4\left(\frac{[\dot{r}_nn]}{n}\right)\epsilon_t^4\right|=o_p(1).
\end{eqnarray}
From (\ref{tau1}) and (\ref{tau2}), we have that
\begin{eqnarray}\label{tau}
\tau^2-\dot{\tau}^2=o_p(1).
\end{eqnarray}In view of (\ref{RS}) and (\ref{tau}), we deduce that $q^{-\frac{1}{2}}\hat{B}_{k,\dot{r}_n}^{\gamma}$ and $q^{-\frac{1}{2}}\dot{B}_{k,\dot{r}_n}^{\gamma}$ have the same asymptotic behavior.
The rest of the proof follows the same arguments as in the proof of
Proposition 2 in Sans\'{o} et al. (2004) and considering
$q^{-\frac{1}{2}}\dot{B}_{k,\dot{r}_n}^{\gamma}$.$\quad\square$\\

\noindent {\bf Proof of Proposition \ref{modtest2}.}\quad Under the alternative hypothesis, the variance can
be written as $g^2(\frac{t}{n})=v(\frac{t}{n})+\alpha
\mathds{1}_{\{t\geq[n\kappa]\}}$, where $[n\kappa]$ is the break
location with $\kappa\in(0,1)$. The function $v(.)$ satisfies a
Lipschitz condition with  $\sup_{r\in(0,1]}|v(r)|<\infty$. Note that
under the alternative hypothesis, the break point is located on the
subsample $[[\dot{r}_nn]+1,[\dot{r}_nn]+q]$, so that there exists
$l\in(0,1)$ such that $[n\kappa]$ can be written as
$[n\kappa]=[\dot{r}_nn]+[lq]+1$. We have
\begin{eqnarray*}
\left|q^{-\frac{1}{2}}B_{k,\dot{r}_n}^{\gamma}\right|&=&q^{-\frac{1}{2}}\left|\frac{\left(C_{k,\dot{r}_n}^{\gamma}-
\frac{k}{q}C_{q,\dot{r}_n}^{\gamma}\right)}
{\sqrt{\eta_{\dot{r}_n}^{\gamma}-\left(q^{-1}C_{q,\dot{r}_n}^{\gamma}\right)^2}}\right|\\
&=&q^{-\frac{1}{2}}\left|C_{k,\dot{r}_n}^{\gamma}-
\frac{k}{q}C_{q,\dot{r}_n}^{\gamma}\right|\times
\left[\hat{\eta}_{\dot{r}_n}^{\gamma}-\left(q^{-1}C_{q,\dot{r}_n}^{\gamma}\right)^2\right]^{-\frac{1}{2}}\\
&=&q^{-\frac{1}{2}}\left|\sum_{t=[\dot{r}_nn]+1}^{[\dot{r}_nn]+k}u_t^2-
\frac{k}{q}\sum_{t=[\dot{r}_nn]+1}^{[\dot{r}_nn]+q}u_t^2\right|\times
\left[q^{-1}\sum_{t=[\dot{r}_nn]+1}^{[\dot{r}_nn]+q}u_t^4-\left(q^{-1}\sum_{t=[\dot{r}_nn]+1}^{[\dot{r}_nn]+q}u_t^2\right)^2\right]^{-\frac{1}{2}}\\
&=&\left|\frac{1}{\sqrt
q}\left[\sum_{t=[\dot{r}_nn]+1}^{[\dot{r}_nn]+k}v\left(\frac{t}{n}\right)\epsilon^2_t-
\frac{k}{q}\sum_{t=[\dot{r}_nn]+1}^{[\dot{r}_nn]+q}v\left(\frac{t}{n}\right)\epsilon^2_t\right]+\frac{\alpha}{\sqrt
q}\left[\sum_{t=[n\kappa]}^{[\dot{r}_nn]+k} \epsilon^2_t-
\frac{k}{q}\sum_{t=[n\kappa]}^{[\dot{r}_nn]+q}
\epsilon^2_t\right]\right|\\
&\times&
\left\{\frac{1}{q}\sum_{t=[\dot{r}_nn]+1}^{[\dot{r}_nn]+q}\left[v\left(\frac{t}{n}\right)+\alpha
\mathds{1}_{\{t\geq[n\kappa]\}}\right]^2\epsilon_t^4
-\left[\frac{1}{q}\sum_{t=[\dot{r}_nn]+1}^{[\dot{r}_nn]+q}\left[v\left(\frac{t}{n}\right)+\alpha  \mathds{1}_{\{t\geq[n\kappa]\}}\right]\epsilon^2_t\right]^2\right\}^{-\frac{1}{2}}\\
&:=&\left|d_1+d_2\right|\times
|d_3-d_4|^{-\frac{1}{2}}:=D_1\times\frac{1}{D_2}.
\end{eqnarray*} From the same arguments used to prove equation (\ref{RS}), it is easy to see that $d_1=O_p(1).$ Let $k=[sq]$ where $s\in(0,1]$, so by applying
the Donsker's functional central limit theorem and the law of large
numbers, we have
\begin{eqnarray*}
&d_2:=&\frac{\alpha}{\sqrt
q}\left[\sum_{t=[\dot{r}_nn]+[lq]+1}^{[\dot{r}_nn]+[sq]}\epsilon^2_t-\frac{[sq]}{q}\sum_{t=[\dot{r}_nn]+[lq]+1}^{[\dot{r}_nn]+q}
\epsilon^2_t\right]\\
&=&\alpha\left[\left(\frac{1}{\sqrt q}\sum_{t=[\dot{r}_nn]+1}^{[\dot{r}_nn]+[sq]}\epsilon^2_t-\frac{1}{\sqrt q}\sum_{t=[\dot{r}_nn]+1}^{[\dot{r}_nn]+[lq]}\epsilon^2_t\right)-\frac{[sq]}{\sqrt q}\left(\frac{1}{q}\sum_{t=[\dot{r}_nn]+1}^{[\dot{r}_nn]+q}
\epsilon^2_t-\frac{1}{q}\sum_{t=[\dot{r}_nn]+1}^{[\dot{r}_nn]+[lq]}\epsilon^2_t\right)\right]\\
&=&\alpha l(s-1)\sqrt{q}+O_p(1).
\end{eqnarray*}Thus,
\begin{eqnarray*}
\sup_{s\in(0,1]} D_1&=&\sup_{s\in(0,1]}\alpha
l(1-s)\sqrt{q}+O_p(1)=M\sqrt{q}+O_p(1).
\end{eqnarray*}
Now let us evaluate the probability limit of $D_2$. Using the same
arguments as for (\ref{tau1}) and (\ref{tau2}), we have
\begin{eqnarray*}
&d_3:=&\frac{1}{q}\sum_{t=[\dot{r}_nn]+1}^{[\dot{r}_nn]+q}\left(v\left(\frac{t}{n}\right)+\alpha
\mathds{1}_{\{t\geq[n\kappa]\}}\right)^2\epsilon^4_t\\&=&\frac{1}{q}\sum_{t=[\dot{r}_nn]+1}^{[\dot{r}_nn]+q}v^2\left(\frac{t}{n}\right)\epsilon^4_t+\frac{\alpha^2}{q}\sum_{t=[\dot{r}_nn]+1}^{[\dot{r}_nn]+q}
\mathds{1}_{\{t\geq [\dot{r}_nn]+[lq]+1\}}\epsilon^4_t+\frac{2\alpha}{q}\sum_{t=[\dot{r}_nn]+1}^{[\dot{r}_nn]+q}\mathds{1}_{\{t\geq [\dot{r}_nn]+[lq]+1\}}v\left(\frac{t}{n}\right)\epsilon^4_t\\
&=&\frac{1}{q}\sum_{t=[\dot{r}_nn]+1}^{[\dot{r}_nn]+q}v^2\left(\frac{t}{n}\right)\epsilon^4_t+\frac{\alpha^2}{q}\sum_{t=[\dot{r}_nn]+1}^{[\dot{r}_nn]+q}\epsilon^4_t-\frac{\alpha^2}{q}\sum_{t=[\dot{r}_nn]+1}^{[\dot{r}_nn]+[lq]}\epsilon^4_t\\
&+&\frac{2\alpha}{q}\sum_{t=[\dot{r}_nn]+1}^{[\dot{r}_nn]+q}v\left(\frac{t}{n}\right)\epsilon^4_t-\frac{2\alpha}{q}\sum_{t=[\dot{r}_nn]+1}^{[\dot{r}_nn]+[lq]}v\left(\frac{t}{n}\right)\epsilon^4_t\\
&=&E(\epsilon_1^4)\left[v^2(\dot{r})+\alpha(1-l)\left(\alpha+2v(\dot{r})\right)\right]+o_p(1).
\end{eqnarray*}
Similarly, it can be shown that
\begin{eqnarray*}
&d_4:=&\left[\frac{1}{q}\sum_{t=[\dot{r}_nn]+1}^{[\dot{r}_nn]+q}v\left(\frac{t}{n}\right)+\alpha
\mathds{1}_{\{t\geq[n\kappa]\}}\epsilon^2_t\right]^2=\left[v(\dot{r})+\alpha(1-l)\right]^2+o_p(1).\\
\end{eqnarray*}Consequently, we can see that $D_2$ is asymptotically constant and finally we have
\begin{eqnarray*}
\sup_k\left|q^{-\frac{1}{2}}\hat{B}_{k,\dot{r}_n}^{\gamma}\right|=M\sqrt{q}+O_p(1).\quad\square
\end{eqnarray*}

The following lemma is used to prove the asymptotic consistency of
polynomial regression estimators described in (\ref{regression1}).
\begin{lem}\label{lemma 1} Suppose that {\bf A1} holds true, then as $q\to\infty$
\begin{equation*}
\hat{\alpha}_{j,r^0}-\alpha_{j,r^0}=o_{p}(1),\quad\mbox{for all}\quad 0<j\leq p.
\end{equation*}
\end{lem}

\noindent {\bf Proof of Lemma \ref{lemma 1}.}\quad The model
(\ref{regression1}) can be expressed in matrix notation as follows:
$U=X\Lambda+\xi,$ where
\begin{center}
$U=\left(
  \begin{array}{c}
    u^{2}_{[\dot{r}_nn]+1} \\
    \vdots \\
    u^{2}_{[\dot{r}_nn]+q} \\
  \end{array}
\right)$, $X=\left(
\begin{array}{cccc}
1 & \frac{[\dot{r}_nn]+1}{n}-r^0 & \ldots &\left (\frac{[\dot{r}_nn]+1}{n}-r^0\right)^p \\
\vdots & \vdots &  & \vdots\\
1 & \frac{[\dot{r}_nn]+q}{n}-r^0 & \ldots &\left (\frac{[\dot{r}_nn]+q}{n}-r^0\right)^p \\
 \end{array}
 \right)$,
$\Lambda=\left(
\begin{array}{c}
\alpha_{0,r^0} \\
\vdots \\
\alpha_{p,r^0} \\
\end{array}
\right)$and $\xi=\left(
  \begin{array}{c}
    \xi_{[\dot{r}_nn]+1} \\
    \vdots \\
    \xi_{[\dot{r}_nn]+q} \\
  \end{array}
\right).$
\end{center}
\noindent The least squares estimate of $\Lambda$ is given by
\begin{eqnarray*}
\hat{\Lambda}&=&(X'X)^{-1}X'U=(X'X)^{-1}X'(X\Lambda+\xi)=\Lambda+(X'X)^{-1}X'\xi,
\end{eqnarray*}so it follows that
\begin{eqnarray*}
\hat{\Lambda}-\Lambda&=&(X'X)^{-1}X'\xi=
\left(\frac{X'X}{q}\right)^{-1}\left(\frac{X'\xi}{q}\right).
\end{eqnarray*}It is clear that
$\left(\frac{X'X}{q}\right)^{-1}=O(1)$. To finish the proof we only
need to show that $\left(\frac{X'\xi}{q}\right)=o_{p}(1).$ By
definition we have
$$\frac{X'\xi}{q}=\left(
\begin{array}{c}
\frac{1}{q}\sum_{t=[\dot{r}_nn]+1}^{[\dot{r}_nn]+q}X_t^0(u_t^{2}-E(u_t^{2})) \\
\frac{1}{q}\sum_{t=[\dot{r}_nn]+1}^{[\dot{r}_nn]+q}X_t^1(u_t^{2}-E(u_t^{2})) \\
\vdots \\
\frac{1}{q}\sum_{t=[\dot{r}_nn]+1}^{[\dot{r}_nn]+q}X_t^q(u_t^{2}-E(u_t^{2})) \\
\end{array}\right),$$ where $X_t^j=(\frac{t}{n}-r^0)^j,
j=0,\cdots,p$ and $\xi_t=u_t^{2}-g^2(\frac{t}{n})$. Note that
$E[\xi_t^2]=E[(u_t^{2}-g^2(\frac{t}{n}))^2]=g^4(\frac{t}{n})[E(\epsilon_t^4-1)]<\infty$.
Thus, by applying Corollary $3.9$ in White (1984), we get
$\frac{1}{q}\sum_{t=[\dot{r}_nn]+1}^{[\dot{r}_nn]+q}\left[u_t^{2}-g^2(\frac{t}{n})\right]=o_p(1),$
which completes the proof.$\quad\square$\\

\noindent{\bf Proof of Proposition \ref{propo4}.}\quad We compare
the statistic defined by (\ref{stat3}) to the statistic defined as
\begin{eqnarray*}\label{stat4}
S_{\dot{r}_n}^{\gamma}=\sup_k|q^{-\frac{1}{2}}B_{k,\dot{r}_n}^{\gamma}|,\quad\mbox{with}\quad
B_{k,\dot{r}_n}^{\gamma}=\frac{C_{k,\dot{r}_n}^{\gamma}-
\frac{k}{q}C_{q,\dot{r}_n}^{\gamma}}
{\sqrt{\eta_{\dot{r}_n}^{\gamma}-(q^{-1}C_{q,\dot{r}_n}^{\gamma})^2}},\quad
k=1,\dots,q,
\end{eqnarray*}where $C_{k,\dot{r}_n}^{\gamma}=\sum_{t=[\dot{r}_nn]+1}^{[\dot{r}_nn]+k}\epsilon_t^2$
and
$\eta_{\dot{r}_n}^{\gamma}=q^{-1}\sum_{t=[\dot{r}_nn]+1}^{[\dot{r}_nn]+q}\epsilon_t^4$.\\
There are two parts of the proof of proposition 4. We study the
numerator and the denominator in (\ref{stat3}) separately. For the
nominator, we have
\begin{eqnarray*}
&&\left|\frac{1}{\sqrt{q}}\left(\bar{C}_{k,\dot{r}_n}^{\gamma}-\frac{k}{q}\bar{C}_{q,\dot{r}_n}^{\gamma}\right)-\frac{1}{\sqrt{q}}\left(C_{k,\dot{r}_n}^{\gamma}-\frac{k}{q}C_{q,\dot{r}_n}^{\gamma}\right)\right|\\
&=&\frac{1}{\sqrt{q}}\left|\sum_{t=[\dot{r}_nn]+1}^{[\dot{r}_nn]+k}\left(\frac{g^2(\frac{t}{n})}{\hat{g}^2(\frac{t}{n})}-1\right)\epsilon_t^2-\frac{k}{q}\sum_{t=[\dot{r}_nn]+1}^{[\dot{r}_nn]+q}\left(\frac{g^2(\frac{t}{n})}{\hat{g}^2(\frac{t}{n})}-1\right)\epsilon_t^2\right|\\
&\leq&\frac{1}{\sqrt{q}}\left|\sum_{t=[\dot{r}_nn]+1}^{[\dot{r}_nn]+k}\left(\frac{g^2(\frac{t}{n})}{\hat{g}^2(\frac{t}{n})}-1\right)\epsilon_t^2\right|+\frac{k}{q^{\frac{3}{2}}}\left|\sum_{t=[\dot{r}_nn]+1}^{[\dot{r}_nn]+q}\left(\frac{g^2(\frac{t}{n})}{\hat{g}^2(\frac{t}{n})}-1\right)\epsilon_t^2\right|\\
&\leq&\sup_{[\dot{r}_nn]+1\leq t\leq [\dot{r}_nn]+q}\left|\frac{g^2(\frac{t}{n})}{\hat{g}^2(\frac{t}{n})}-1\right|\left(\frac{1}{\sqrt{q}}\left|\sum_{t=[\dot{r}_nn]+1}^{[\dot{r}_nn]+k}\epsilon_t^2\right|+\frac{k}{q^{\frac{3}{2}}}\left|\sum_{t=[\dot{r}_nn]+1}^{[\dot{r}_nn]+q}\epsilon_t^2\right|\right)\\
&\leq&\sup_{[\dot{r}_nn]+1\leq t\leq
[\dot{r}_nn]+q}\left|\frac{g^2(\frac{t}{n})-\hat{g}^2(\frac{t}{n})}{\hat{g}^2(\frac{t}{n})}\right|\left(\frac{1}{\sqrt{q}}\left|\sum_{t=[\dot{r}_nn]+1}^{[\dot{r}_nn]+k}(\epsilon_t^2-1)\right|+\frac{k}{q^{\frac{3}{2}}}\left|\sum_{t=[\dot{r}_nn]+1}^{[\dot{r}_nn]+q}(\epsilon_t^2-1)\right|+\frac{2k}{\sqrt{q}}\right).
\end{eqnarray*}We consider a large enough $n$ such that
$\hat{g}^2(\frac{t}{n})>c>0$. Then
\begin{eqnarray}\label{eqn}
\nonumber&&\sup_{[\dot{r}_nn]+1\leq
t\leq[\dot{r}_nn]+q}\left|\frac{g^2(\frac{t}{n})-\hat{g}^2(\frac{t}{n})}{\hat{g}^2(\frac{t}{n})}\right|\\\nonumber
&\leq&\frac{1}{c}\sup_{[\dot{r}_nn]+1\leq
t\leq[\dot{r}_nn]+q}\left|g^2\left(\frac{t}{n}\right)-\hat{g}^2\left(\frac{t}{n}\right)\right|\\\nonumber
&\leq&\frac{1}{c}\sup_{[\dot{r}_nn]+1\leq
t\leq[\dot{r}_nn]+q}\left|\sum_{i=0}^p(\alpha_{i,r^0}-\hat{\alpha}_{i,r^0})\left(\frac{t}{n}-r^0\right)^i\right|+o\left[\left(\frac{t}{n}-r^0\right)^p\right]\\\nonumber
&\leq&\frac{1}{c}\sup_{[\dot{r}_nn]+1\leq
t\leq[\dot{r}_nn]+q}\sum_{i=0}^p\left|\alpha_{i,r^0}-\hat{\alpha}_{i,r^0}\right|\left|\frac{t}{n}-r^0\right|^i+o\left(n^{p(\gamma-1)}\right)\\\nonumber
&\leq&\frac{1}{c}\frac{q}{2n}\sum_{i=0}^p\left|\alpha_{i,r^0}-\hat{\alpha}_{i,r^0}\right|+o\left(n^{p(\gamma-1)}\right)\\
&=&o(n^{(\gamma-1)})+o(n^{p(\gamma-1)}),
\end{eqnarray} where the last equality follows from Lemma \ref{lemma 1}. Therefore, it follows from (\ref{eqn}), the Donsker
Theorem's and the law of large numbers that
\begin{eqnarray}\label{eq5}
\left|\frac{1}{\sqrt{q}}\left(\bar{C}_{k,\dot{r}_n}^{\gamma}-\frac{k}{q}\bar{C}_{q,\dot{r}_n}^{\gamma}\right)-
\frac{1}{\sqrt{q}}\left(C_{k,\dot{r}_n}^{\gamma}-\frac{k}{q}C_{q,\dot{r}_n}^{\gamma}\right)\right|
&=&o_p(1),
\end{eqnarray}for all $0<\gamma<\frac{2}{3}$.\\ For the denominator we introduce
\begin{eqnarray*}
\bar{\tau}^2
=\bar{\eta}_{\dot{r}_n}^{\gamma}-(q^{-1}\bar{C}_{q,\dot{r}_n}^{\gamma})^2=\frac{1}{q}\sum_{t=[\dot{r}_nn]+1}^{[\dot{r}_nn]+q}\frac{g^4(\frac{t}{n})}{\hat{g}^4(\frac{t}{n})}\epsilon_t^4-\left(
\frac{1}{q}\sum_{t=[\dot{r}_nn]+1}^{[\dot{r}_nn]+q}\frac{g^2(\frac{t}{n})}{\hat{g}^2(\frac{t}{n})}\epsilon_t^2\right)^2
\end{eqnarray*}and
\begin{eqnarray*}
\tau^2=\eta_{\dot{r}_n}^{\gamma}-(q^{-1}C_{q,\dot{r}_n}^{\gamma})^2=\frac{1}{q}\sum_{t=[\dot{r}_nn]+1}^{[\dot{r}_nn]+q}\epsilon_t^4-\left(
\frac{1}{q}\sum_{t=[\dot{r}_nn]+1}^{[\dot{r}_nn]+q}\epsilon_t^2\right)^2.
\end{eqnarray*}We have
\begin{eqnarray*}
&&\left|\frac{1}{q}\sum_{t=[\dot{r}_nn]+1}^{[\dot{r}_nn]+q}\frac{g^2(\frac{t}{n})}{\hat{g}^2(\frac{t}{n})}\epsilon_t^2-
\frac{1}{q}\sum_{t=[\dot{r}_nn]+1}^{[\dot{r}_nn]+q}\epsilon_t^2\right|\leq \frac{1}{q}\sum_{t=[\dot{r}_nn]+1}^{[\dot{r}_nn]+q}\left|\frac{g^2(\frac{t}{n})}{\hat{g}^2(\frac{t}{n})}-1\right|\epsilon_t^2\\
&\leq&\sup_{[\dot{r}_nn]+1\leq
t\leq[\dot{r}_nn]+q}\left|\frac{g^2(\frac{t}{n})-\hat{g}^2(\frac{t}{n})}{\hat{g}^2(\frac{t}{n})}\right|\times\frac{1}{q}\sum_{t=[\dot{r}_nn]+1}^{[\dot{r}_nn]+q}\epsilon_t^2.
\end{eqnarray*}Using (\ref{eqn}) and the law of large numbers, we get
\begin{eqnarray*}
\left|\frac{1}{q}\sum_{t=[\dot{r}_nn]+1}^{[\dot{r}_nn]+q}\frac{g^2(\frac{t}{n})}{\hat{g}^2(\frac{t}{n})}\epsilon_t^2-
\frac{1}{q}\sum_{t=[\dot{r}_nn]+1}^{[\dot{r}_nn]+q}\epsilon_t^2\right|=o_p(1).
\end{eqnarray*}Similarly, we write
\begin{eqnarray*}
\left|\frac{1}{q}\sum_{t=[\dot{r}_nn]+1}^{[\dot{r}_nn]+q}\frac{g^4(\frac{t}{n})}{\hat{g}^4(\frac{t}{n})}\epsilon_t^4-
\frac{1}{q}\sum_{t=[\dot{r}_nn]+1}^{[\dot{r}_nn]+q}\epsilon_t^4\right|=o_p(1),
\end{eqnarray*} which implies that
\begin{eqnarray}\label{res2}
\bar{\tau}^2-\tau^2=o_p(1).
\end{eqnarray}
As a result, from (\ref{eq5}) et (\ref{res2}), we deduce that
$$|q^{-\frac{1}{2}}\bar{B}_{k,\dot{r}_n}^{\gamma}-q^{-\frac{1}{2}}B_{k,\dot{r}_n}^{\gamma}|=o_p(1),$$ and that $q^{-\frac{1}{2}}\bar{B}_{k,\dot{r}_n}^{\gamma}$
and $q^{-\frac{1}{2}}B_{k,\dot{r}_n}^{\gamma}$ have the same
asymptotic behavior. The rest of the proof follows the same
arguments as in the proof of Proposition 2 in Sans\'{o} et al. (2004)
and considering $q^{-\frac{1}{2}}B_{k,\dot{r}_n}^{\gamma}$.$\quad\square$\\

\end{document}